\documentclass{ws-procs9x6}

\usepackage{xspace}

\def\cref#1{Chapt.\,\ref{#1}}
\def\Cref#1{Chapter~\ref{#1}}

\def\fref#1{Fig.\,\ref{#1}}

\def\eref#1{(\ref{#1})}

\def\rref#1{Ref.\,\cite{#1}}

\def\lleft{\textit{left}}
\def\rright{\textit{right}}
\def\LLeft{\textit{Left}}
\def\RRight{\textit{Right}}
\def\ttop{\textit{top}}
\def\bbottom{\textit{bottom}}

\def\1{\footnotemark[1]}
\def\and{\& }
\def\Cerenkov{\v{C}erenkov\xspace}

\def\deg{$^\circ$\xspace}

\def\gcm2{g/cm$^2$\xspace}

\def\Xmax{$X_{max}$\xspace}
\def\vxB{$\vec{v}\times\vec{B}$\xspace}
\def\Lstar{LOPES$^{\rm STAR}$\xspace}

\begin{document}

\title{MEASUREMENT OF RADIO EMISSION FROM EXTENSIVE AIR SHOWERS}

\author{\sc J\"org R. H\"orandel}

\address{Radboud University Nijmegen, Department of Astrophysics,\\
P.O. Box 9010, 6500 GL Nijmegen, The Netherlands\\
E-mail: j.horandel@astro.ru.nl, http://particle.astro.kun.nl}

\begin{abstract}
A new promising development in astroparticle physics is to measure the radio
emission from extensive air showers. The particles in the cascade emit
synchrotron radiation (30 - 90 MHz) which is detected with arrays of dipole
antennas.  Recent experimental efforts are discussed.
\end{abstract}


\bodymatter

\section{Introduction}

An intense branch of astroparticle physics is the study of high-energy cosmic
rays to reveal their origin, as well as their acceleration and propagation
mechanisms \cite{behreview,cospar06,wuerzburg}. At energies exceeding
$10^{14}$~eV cosmic rays are usually studied by indirect measurements --- the
investigation of extensive air showers initiated by cosmic particles in the
atmosphere. Different techniques are applied, like the measurements of particle
densities and energies at ground level, or the observation of \Cerenkov and
fluorescence light. An alternative technique has been recently revitalized ---
the detection of radio emission (at tens of MHz) from extensive air showers at
energies exceeding $10^{16}$~eV.

Radio emission from air showers was experimentally discovered in 1965 at a
frequency of 44 MHz \cite{jelleynature}. The early activities in the 1960s and
1970s are summarized in \rref{allanrev}.  Only recently, fast analog-to-digital
converters and modern computer technology made a clear detection of radio
emission from air showers possible.  LOPES, a LOFAR Prototype Station had shown
that radio emission from air showers can be detected even in an environment
with relatively strong radio frequency interference (RFI) \cite{radionature}.
Further investigations of the radio emission followed with LOPES
\cite{badearadio,petrovicinclined,buitinkthunder,niglfreq,nigldirect} and the
CODALEMA experiment \cite{codalema,Ardouin:2006nb}, paving the way for this new
detection technique \cite{Falcke:2008qk}. 

Most likely, the dominant emission mechanism of the radio waves in the
atmosphere is synchrotron radiation due to the deflection of charged particles
in the Earth magnetic field (geosynchrotron radiation)
\cite{huege2003,huege2005,huegefalcke}.  In the frequency range of interest
($30-90$~MHz) the wavelength of the radiation is large compared to the size of
the emission region: the typical thickness of the shower disc is about 1 to 2~m
only. Thus, coherent emission is expected, which yields relatively strong
signals (of the order of a few $\mu$V/m MHz) at ground level.

Goal of the present activities is to further push the development of radio
detection to become a new, independent way to measure the properties of air
showers. Ultimate goal is to derive information about the primary,
shower-inducing particle from the measurements, such as the energy and mass of
the particle as well as the particle direction and point of incidence.  An
advantage of radio detection, e.g.\ with respect to the fluorescence technique,
is that showers can be observed with almost 100\% duty cycle.

The {\sl arrival direction} can be inferred from the arrival time of the
wavefront at the antennas.
The measured radio signal at a certain distance from the shower axis
is a good estimator for the number of particles at shower
maximum, being almost independent of the primary particle type
\cite{huege2008}. This can be used to determine the shower {\sl energy}.
The curvature of the shower front has been investigated
\cite{lafebrecurv,icrc09-lafebre}. It could be shown that the radius of
curvature measured at ground level is related to the distance to the shower
maximum.  The depth of the shower maximum in the atmosphere \Xmax is an
important observable to determine the mass of the primary particle.  The study
indicates a resolution of \Xmax from radio observations of order of
$30-40$~\gcm2, i.e.\ comparable to the accuracy of present fluorescence
detectors.

\section{Experimental Set-Ups}

\subsection{LOPES}
The LOPES experiment registers radio signals in the frequency range from 40 to
80~MHz \cite{lopesspie}. In this band are few strong man made radio
transmitters only, the emission from air showers is still strong (it decreases
with frequency), and background emission from the Galactic plane is still low.
An active short dipole has been chosen as antenna.  An inverted V-shaped dipole
is positioned about 1/4 of the shortest wavelength above an aluminum ground
plate. In this way a broad directional beam pattern is obtained.  LOPES
comprises 30 antennas \cite{nehlspune} located on site of the KASCADE-Grande
experiment \cite{kascadenim,grande}.  The LOPES data acquisition is triggered
by large air showers registered with KASCADE-Grande. The latter measures the
showers simultaneously to LOPES and delivers precise information on the shower
parameters, such as shower energy as well as position and inclination of the
shower axis.  

\subsection{CODALEMA}

The CODALEMA experiment is set up at the Nancay radio observatory
\cite{codalema,Ravel:2003mi}.  In the first phase six conic logarithmic
periodic antennas of the Decametric array have been used.  The present set-up
comprises new antennas, developed to optimize simplicity, costs, and
performance \cite{Ardouin:2007zz}. The broadband antennas (100~kHz -- 220~MHz)
are based on a fat active dipole concept. They are simple dipoles ($2\times
0.6$~m long) placed 1~m above ground.  21 dipoles are oriented in east-west
polarization direction and three dipoles in north-south direction.  The set-up
is completed by 17 plastic scintillation counters to measure the properties of
air showers.

\subsection{LOFAR}

The Low Frequency Array (LOFAR) is a new digital radio observatory, presently
under construction \cite{lofar,lofar-isvhecri08}.  An objective of LOFAR is the
detection of radio emission from particle cascades, originating from extremely
high-energy particles from outer space.  Two main lines of research are
followed: (i) the measurement of radio emission from extensive air showers,
generated by interactions of high-energy cosmic rays in the atmosphere
\cite{icrc09-horneffer} and (ii) the detection of radio emission of particle
cascades in the Moon, originating from ultra high-energy neutrinos and cosmic
rays interacting with the lunar surface \cite{icrc09-singh}.

More than 40 stations with fields of relatively simple antennas will work
together as digital radio interferometer. The antenna fields are distributed
over several countries in Europe with a dense core in the Netherlands. The
latter will have at least 18 stations in an area measuring roughly
$2\times3$~km$^2$. Each station will comprise 96 low band antennas, simple
inverted V-shaped dipoles (like the LOPES antennas), operating in the frequency
range from 30 to 80~MHz.  Each antenna will have a dipole oriented in
north-south and east-west directions, respectively. In addition, fields
\footnote{The fields comprise 48 antennas in the Dutch stations and 96 in the
European ones.} of high-band antennas will cover the frequency range from 110
to 240~MHz.  For air shower observations the signals from the low band antennas
are digitized and stored in a ring buffer (transient buffer board, TBB).  For
valid triggers the data are send to a central processing facility, based on an
IBM Blue Gene supercomputer.  First data from LOFAR are expected in early 2010.

A small air shower array will be installed in the LOFAR core.  It comprises 20
scintillator counters, each with an area of about 1~m$^2$, the detectors have
been used previously in the KASCADE experiment \cite{kascadenim}. The array
will provide basic air shower information.  It may also serve as trigger for
the radio antennas.

\subsection{AERA -- Pierre Auger Observatory}

Radio emission from air showers at the Pierre Auger Observatory will be
measured with the Auger Engineering Radio Array (AERA) \cite{icrc09-vdberg}.
AERA is co-located with the infill array of the Auger Observatory and in
the field of view of the high-altitude fluorescence telescopes (HEAT)
\cite{icrc09-kleifges}. This unique set-up will allow to register air showers
simultaneously with three independent detection methods: radio waves,
fluorescence light, and particle detection in water \Cerenkov detectors.  AERA
will comprise about 150 antennas located on an area of about 20~km$^2$ and is
designed to cover the energy range from $10^{17}$ to $10^{19}$~eV.  During the
first phase (Spring 2010) 24 logarithmic periodic dipole antennas will be
deployed and put into operation.

Presently, different antenna designs are tested and optimized at the Auger
site.  Among them are logarithmic periodic dipole antennas \cite{fliescher2009}
and simple dipoles, as in the CODALEMA experiment \cite{revenu2009}.  A ring
antenna design and FPGA-based hardware for a self trigger algorithm are
developed within \Lstar and at the Auger site
\cite{gemmekearena,icrc09-kroemer,icrc09-schmidt}.

\subsection{Calibration}

\begin{figure}[t]
 \begin{minipage}[b]{0.46\textwidth}
 \epsfig{file=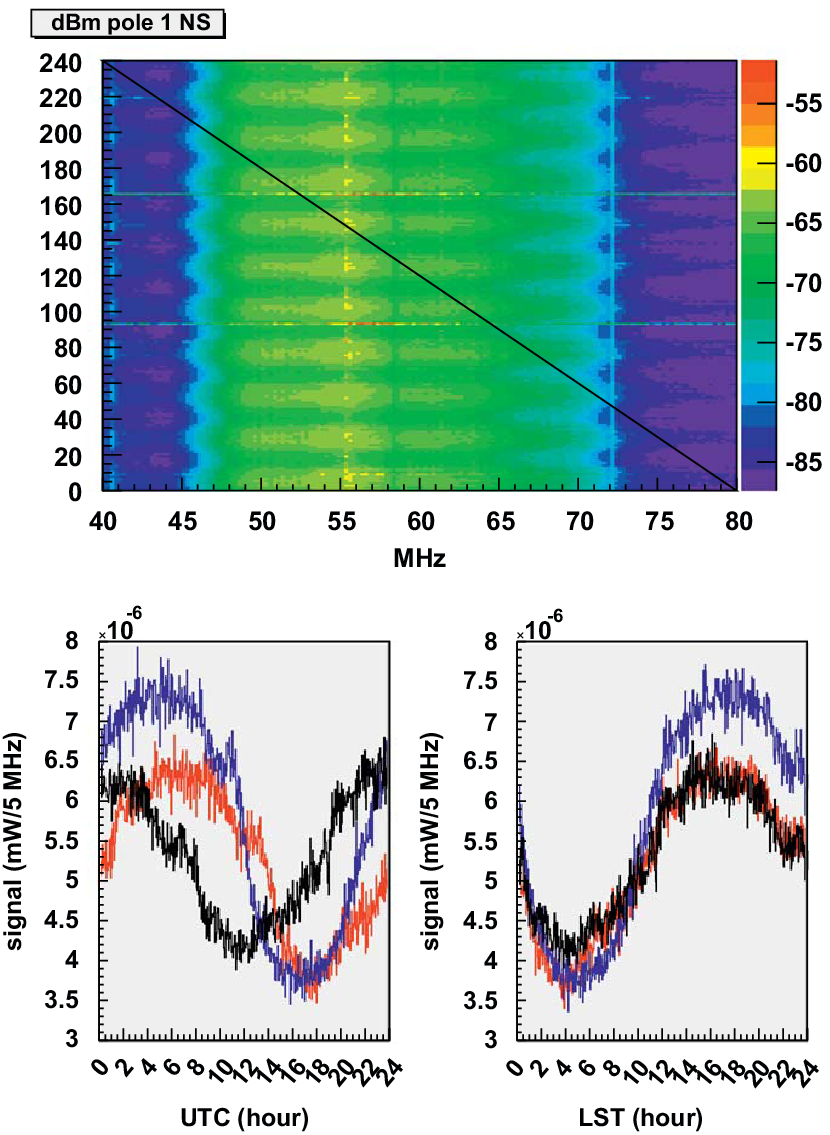,clip=,
         bbllx=115, bblly=0, bburx=237, bbury=160, width=\textwidth}
 \caption{Radio background measured at the Auger Observatory as function of
          siderial time \cite{coppens2009}.\label{siderial}}
 \end{minipage}\hspace{\fill}
 \begin{minipage}[b]{0.52\textwidth}
 \epsfig{file=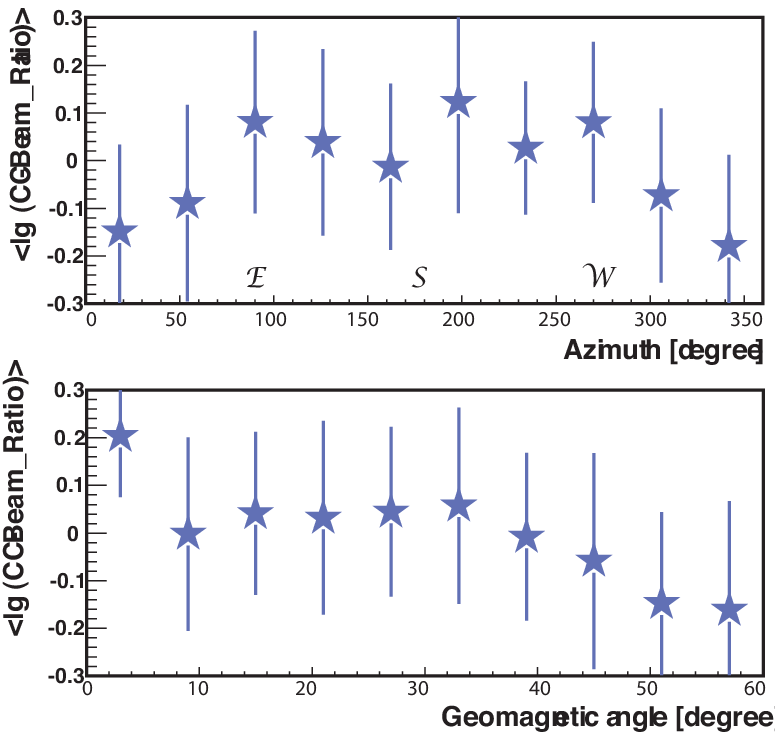,width=\textwidth}
 \caption{Ratio of the measured signal strength of the north-south and
	  east-west polarization components as a function of azimuth angle
	  (top) and geomagnetic angle (bottom) as measured by LOPES
          \cite{icrc09-isar}.
\label{pol}}	  
 \end{minipage} 
\end{figure}

The antenna gain pattern is typically obtained from numerical
calculations.
Two methods are available for the absolute calibration of the electronics.
(i) In LOPES all antennas, including the complete analog electronics chain,
have been individually calibrated with a reference radio source
\cite{Nehls:2008ix}.  
(ii) An independent way is to use the galactic radio background as calibration
source.  The expected strength of the galactic radio signal can be calculated
as function of siderial time as well as day of the year and can be used as
reference.  This technique is presently explored with AERA \cite{coppens2009}.
The measured radio signal as function of siderial time is depicted in
\fref{siderial}. The colors represent measurements at different days of the
year. The siderial variations are clearly visible, illustrating the good
sensitivity of the system.

\section{Radio Signal and Shower Parameters}

The properties of the showers registered by LOPES are simultaneously measured
by the KASCADE-Grande air shower experiment.
An empirical relation has been found to express the expected east-west
component of the field strength at a distance $R$ from the shower axis as a
function of shower parameters \cite{horneffer-merida}
\begin{eqnarray} \label{lopesfun}
\epsilon=(11\pm1) \left[ (1.16\pm0.025)-\cos \alpha \right] \cos \Theta 
   \exp\left(\frac{-R}{(236\pm81)~{\rm m}}\right) \\
     \left(\frac{E_0}{10^{17}~{\rm eV}}\right)^{(0.95\pm0.04)}
     \left[\frac{\mu{\rm V}}{\rm m~MHz}\right]. \nonumber
\end{eqnarray}
$\alpha$ is the angle between the shower axis and the direction of the Earth
magnetic field (geomagnetic angle), $\Theta$ the zenith angle of the shower,
and $E_0$ the energy of the shower inducing primary particle.  It is
interesting to note the absolute values of some parameters: the exponential
fall-off has a characteristic length of about 240~m, much larger than the
classical Moli\`ere radius of electrons in air ($\approx80$~m). This indicates
that the lateral distribution of the radio component is flatter as compared to
the electromagnetic component, an important fact to build large-scale radio
arrays with an economic antenna density.  The measured radio signal is almost
directly proportional to the shower energy ($\propto E_0^{\approx1}$). Such a
behavior is expected for a coherent emission of the radio waves in the air
showers.  This calibration of the measured radio signal is a first important
step towards the application of the radio detection as independent method to
register extensive air showers.

Recently, the lateral distribution of the measured radio signals has been
investigated in more detail with LOPES \cite{icrc09-nehls,jrhicatpp}.  It is
found that some air showers exhibit a relatively flat lateral distribution.
Most likely, such events are registered at a small distance to the shower axis
for showers with large zenith angles.

Also the CODALEMA group has parameterized the expected radio signals as
function of shower parameters \cite{riviereprivate}:
\begin{eqnarray}\label{codalemafun}
 \epsilon=2.3\frac{\mu \mbox{V}}{\mbox{m~MHz}} ~
  \frac{\Delta\nu}{\mbox{MHz}} ~
  \frac{-\vec v\times\vec B}{|\vec v||\vec B|}
  \frac{|\vec B|}{47~\mu\mbox{T}} ~ \cos\Theta ~ 
  \exp\left(\frac{-R}{D_0(\Theta,\nu)}\right) \\
  \frac{E_0}{10^{17}~\mbox{eV}}
  \left[\frac{\mu\mbox{V}}{\mbox{m}}\right]  \nonumber .
\end{eqnarray}
With the parameters $\Delta\nu$ the frequency interval, energy $E_0$ and zenith
angle $\Theta$ of the shower, $\vec v$ the direction of the shower axis,
$\vec B$ the Earth magnetic field, $R$ the distance to the shower axis, and
$D_0(\Theta,\nu)$ an empirical function.
\footnote{The angle between $\vec v$ and $\vec B$ in \eref{codalemafun} is
$\alpha$ in \eref{lopesfun}. The $\alpha$ dependence in \eref{lopesfun} and
\eref{codalemafun} is $(\approx1)-\cos\alpha$ and $|\vec v\times\vec
B|/(|\vec v||\vec B|)=\sin\alpha$, respectively.  In the interval
$[0,\pi/2]$, both expressions yield similar values, which are compatible with
the measurements, given the present accuracy of the data.}

The frequency spectra in air showers have been investigated with both, the
LOPES \cite{niglfreq} and CODALEMA \cite{Ardouin:2006nb} experiments, yielding
similar results.  Due to the relative small frequency range covered
($\approx40-80$~MHz) it can not be decided yet whether the data are described
better by a power law or an exponential function.

\section{Polarization and Asymmetries}

The investigation of the polarization properties and of asymmetries in the
arrival direction is the key to confirm the geosynchrotron mechanism and is
currently the hottest topic discussed by CODALEMA and LOPES.  
The field strength produced near the motion
axis of the particle is expected to be 
approximately proportional to $\vec v\times \vec B$
and the radiation is expected to be linearly polarized
\cite{Riviere:2009pr,Ardouin:2009zp}.  The E-W and N-S polarization components
are obtained by the projection of the $\vec v\times\vec B$ vector on the
corresponding axis. Strong radio signals are expected from showers with arrival
directions being about perpendicular to the magnetic field in the respective
polarization direction.

\begin{figure}[t]
 \epsfig{file=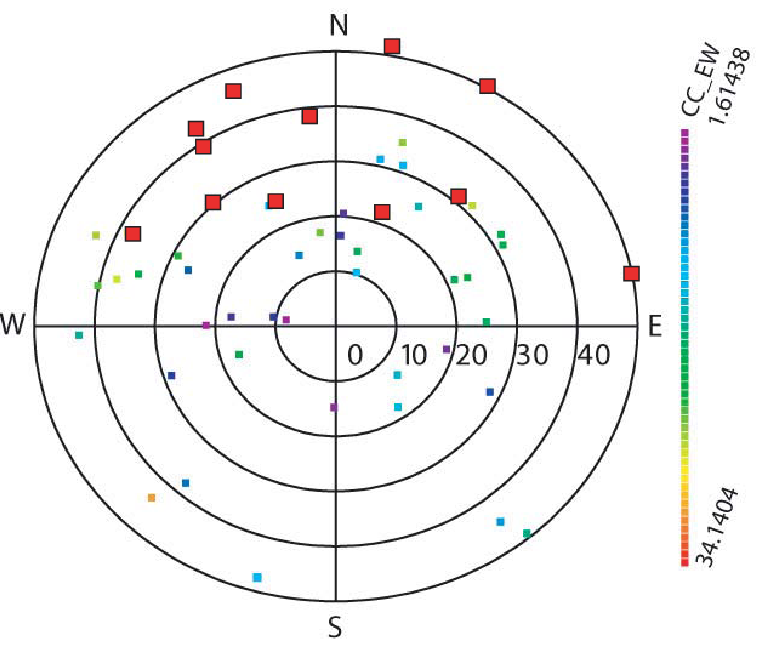,width=0.49\textwidth}\hspace*{\fill}
 \epsfig{file=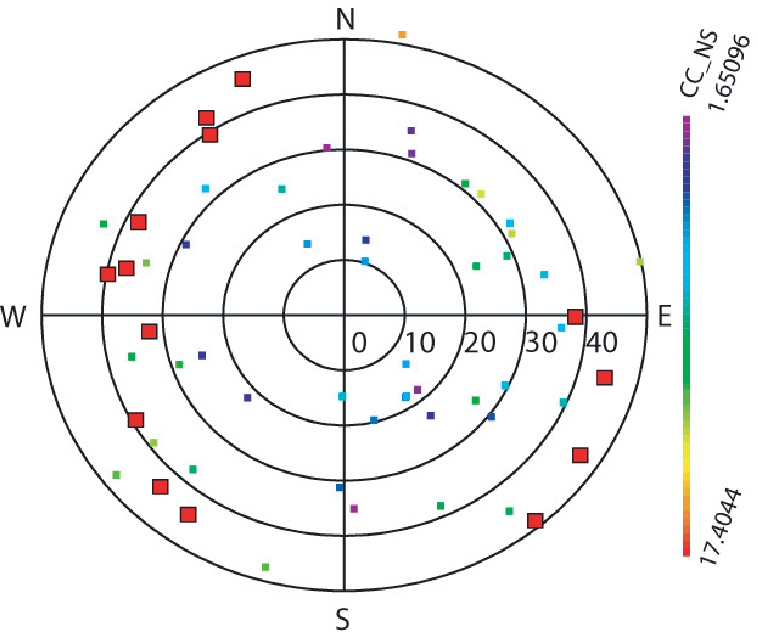,width=0.49\textwidth}
 \caption{Sky map of air showers detected in east-west (left) and north-south
	  (right) polarization direction with LOPES \cite{isar-nim2009}. The
	  boxes mark the arrival direction of the air showers with the
          strongest radio signal.}
 \label{polsky}	  
\end{figure}

\begin{figure}[t]
 \epsfig{file=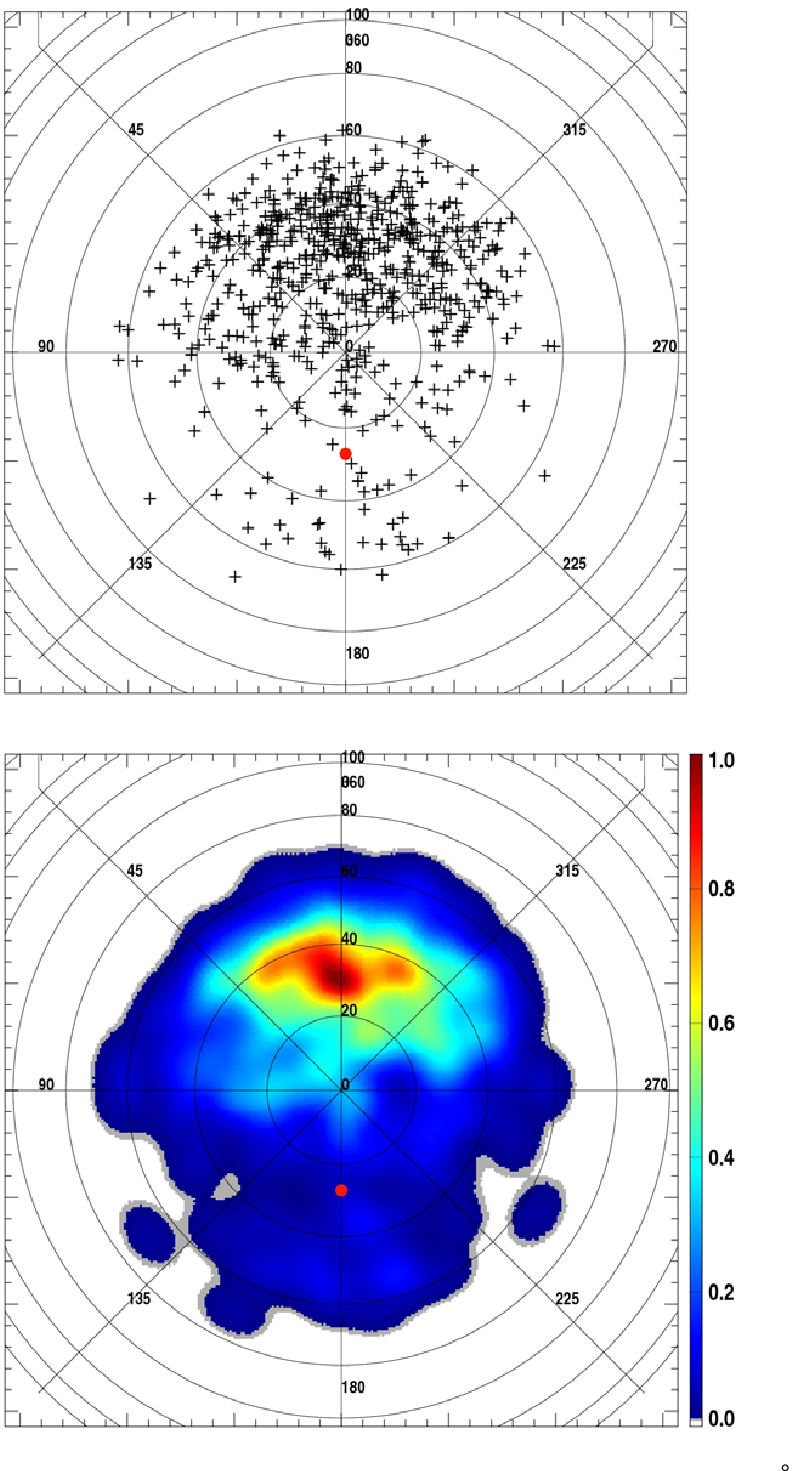,width=0.49\textwidth, clip=,
         bbllx=0, bblly=0, bburx=220, bbury=210}\hspace{\fill}
 \epsfig{file=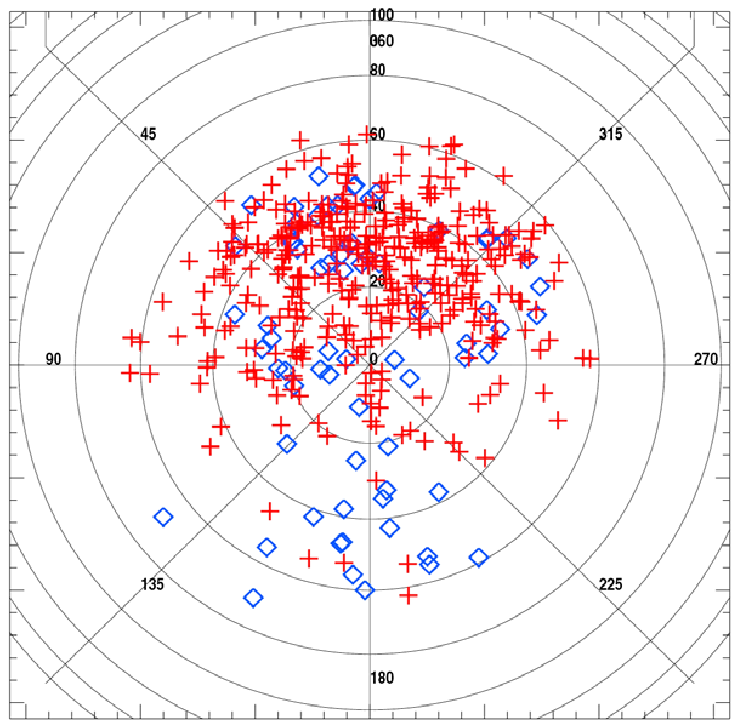,width=0.50\textwidth}
 \caption{Sky maps of events recorded with CODALEMA in E-W polarization
          direction \cite{Ardouin:2009zp}.
          \LLeft: arrival direction of showers.
          \RRight: polarity of measured signal.
          \label{codalemasky}}
\end{figure}

Since the Earth magnetic field is inclined \footnote{Elevation angles of the
magnetic field are LOPES: 65\deg, CODALEMA: 63\deg, Auger: -35\deg.} we expect
an asymmetry in the observed signal strength on the sky.
Sky maps of the E-W component as measured by the different experiments
are shown in \fref{polsky} (\lleft) \cite{isar-nim2009}, \fref{codalemasky}
(\lleft) \cite{Ardouin:2009zp}, and \fref{raugersky} \cite{revenu2009}.  The
color code in \fref{polsky} represents the measured field strengths.  The boxes
indicate the directions of the events with the strongest signals.  LOPES and
CODALEMA clearly see more strong events from the north, while more strong
events are registered from the south in the southern hemisphere.  A similar
effect is observed for the N-S component. In this case, strong events arrive
from the east and the west \cite{isar-nim2009,Riviere:2009pr}, see
\fref{polsky} (\rright).

The ratio of the measured field strength in N-S to E-W polarization directions
as measured by LOPES is depicted as function of azimuth (\ttop) and geomagnetic
angle (\bbottom) in \fref{pol} \cite{isar-nim2009,icrc09-isar}.  Mean values
and the spread of the distributions are shown.  A correlation between the
plotted quantities can be recognized.  The ratio of the north-south divided by
the east-west component of the \vxB vector of the measured showers exhibits the
same behavior as a function of the azimuth angle and geomagnetic angle.  

In addition, the CODALEMA group investigated the polarity of the fields
measured at ground \cite{Riviere:2009pr}. Also in this observable clear
asymmetries are measured. Showers with a positive electric field arrive mainly
from the north and the east, while showers which induce a negative field
preferably come from southern and western directions.  Simulations of the radio
emission by the CODALEMA group show the same behavior as the measured data for
both polarization directions, for the arrival direction of the showers as well
as for the polarity \cite{Riviere:2009pr}.

The recent measurements and simulations conducted at various sites strongly
support an emission process in air showers being proportional to $\vec v\times
\vec B$ and they support geosynchrotron radiation as dominant process.

\section{Thunderstorm Events}

During thunderstorms there are strong (static) electric fields in the
atmosphere which severely influence the radio emission of air showers. The
study of radio emission during thunderstorms has a two-fold goal: (i) the study
of the thunderstorms itself and (ii) the (early) detection of thunderstorm
conditions to prevent false air shower reconstruction during thunderstorms.

\begin{figure}[t]
 \begin{minipage}[b]{0.49\textwidth}
 \epsfig{file=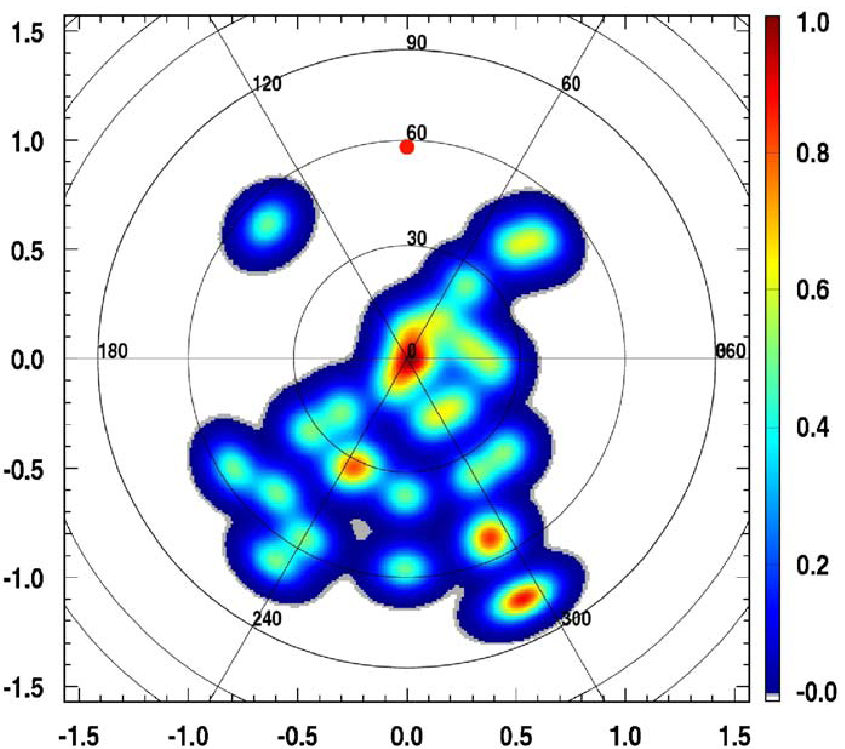,width=\textwidth}
 \caption{Sky map of radio events detected at the Auger Observatory 
          in E-W polarization direction
          \cite{revenu2009}.  \label{raugersky}}
 \end{minipage}\hspace{\fill}
 \begin{minipage}[b]{0.49\textwidth}
 \epsfig{file=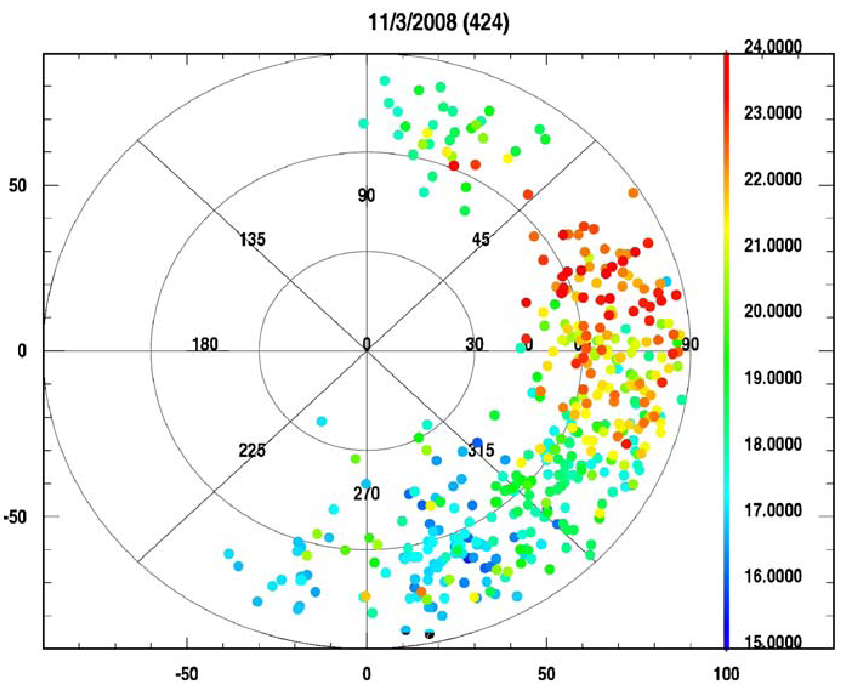,width=\textwidth}
 \caption{Sky map of radio events detected during a thunderstorm at the Pierre
	  Auger Observatory. The colors represent the observation time
          \cite{revenu2009}.
          \label{raugerthunder}}
 \end{minipage}
\end{figure}

As an example, a thunderstorm observed at the Auger Observatory is shown in
\fref{raugerthunder} \cite{revenu2009}. It can be recognized how the
thunderstorm moves on the sky as function of time.

Quantitative investigations of the radio signals during thunderstorms have been
conducted with LOPES data \cite{buitinkthunder,icrc09-ender,jrhicatpp}.  The
electric fields in the atmosphere yield to an acceleration of charged
particles, which in turn yields to an amplification of the radio emission from
air showers during thunder storm conditions.

\section{Outlook} \label{outsec}

The investigations of radio signals from air showers with LOPES and CODALEMA
are the basis for the application of these technique in large-scale
experiments. The next step is to utilize the radio detection technique in large
arrays, comprising several hundreds of antennas.
The radio detection of air showers is a fast growing sub-discipline in
astroparticle physics. With the new experiments starting operation, exciting
results are expected in the next few years.

\section*{Acknowledgments}
I would like to thank the conference organizers for their invitation to give
this review in the inspiring environment of Lago di Como and the Villa Olmo.
I'm thankful to my colleagues from LOPES, LOFAR, and the Pierre Auger
Observatory for fruitful discussions.


\end{document}